\documentstyle[12pt]{article}
\newcommand{\beq}{\begin{equation}\label}
\newcommand{\eeq}{\end{equation}}
\newcommand{\p}{\partial}
\begin{document}
\begin{flushright}
USITP-96-01\\
January 4, 1996
\end{flushright}

\bigskip
\Large
\begin{center}
{\bf A possible description of the quantum numbers in a hadronic 
string model}
\bigskip
\normalsize

 V.A. Kudryavtsev
{\footnote{e-mail: Kudryav@thd.pnpi.spb.ru}}\\ 
{\it Petersburg Nuclear Physics Institute\\
Gatchina, 188350 St. Petersburg, Russia}\\
\bigskip
G. Weidl{\footnote{e-mail:galia@vanosf.physto.se}}\\
{\it Institute of Theoretical Physics\\
 University of Stockholm, \,\, Box 6730\\
 S-113 85 Stockholm, Sweden}
\end{center}
\normalsize
\begin{quote}
{\bf ABSTRACT:} We consider a critical composite superconformal string 
model to desribe hadronic interactions. We present a new approach of 
introducing hadronic quantum numbers in the scattering amplitudes. 
The physical states carry the quantum numbers and form a  common system 
of eigenfunctions of the operators in this string model. We give 
explicit constructions of the quantum number operators. 
\end{quote}

\newpage
\section{Introduction}

In 1967  the Regge hadronic resonances have been discovered.
The experimental data gave evidence of
a typical relation between the spins 
${\cal I}$ and the square of the masses $M$
of these strongly interacting particles \cite{collins}.
It turned out, that the resonances form the so-called
Regge trajectories
\beq{rt}
{\cal I}(M^2) = \alpha_0 + \alpha'M^2 -n\ .
\eeq
For $n=0,1,2,...$ one finds a family of 
parallel linear daughter trajectories. The constants
$\alpha_0$ and $\alpha'$ denote the intercept and the Regge slope. 
The leading $\rho$-meson trajectory corresponds to $n=0.$

So far QCD has not been able to explain this phenomenon.
On the other hand in the 1970's it was observed, that
in string theory 
the mass shell condition 
$$(L_0 - a)|\rm phys>=0$$ 
generates a mass spectrum with parallel 
trajectories. The intercept $a$ depends on the 
cancellation condition of the 
conformal anomaly in the respective theory.
Thus {\it hadronic strings} became a natural candidate
for describing the Regge spectrum phenomena. However such
efforts failed because of the necessary appearance of 
massless states of spin one and two,
which do not correspond to physical hadronic resonances.

While these massless states gave rise to
treat string theory as a fundamental theory
of all interactions including gravity at Planck energy scale,
the discrepance in hadronic string theory at typical strong
interaction energy scales of $E \sim 1$ GeV remained unresolved
for a long time.
In \cite{57}, \cite{k} one of the authors suggested a
new critical composite superconformal hadronic string.
It consists of the Neveu-Schwarz (NS) superstring and
a fermionic superconformal string \cite{vak}. The
NS field components are associated with the space-time
degrees of freedom, while the fermionic sector carries
the internal degrees of freedom, namely the hadronic 
quantum numbers. In accordance with the conformal anomaly
cancelation in this model we have to impose new gauge constraints on
the physical states. This allows to eliminate the problematic
massless states from the string mass spectrum,
which now becomes compatible with the Regge resonance spectrum 
\cite{57}.

In this paper we present a possible explicit construction of the 
operators, corresponding to the hadronic quantum numbers
spin, isospin, electric charge, hyper-charge,
baryon charge, strangeness, charm, beauty and 
top. We give the structure of the hadronic wave functions.

In section 2 we review the main features of the composite string
model and the structure of its superconformal generator. We
discuss the cancellation of the anomaly  achieved in $D=10$  or $D=4$
space-time  dimensions  and $D'$ internal fermionic degrees of freedom.
Then additional $\frac{D'}{6}$ or  $\frac{D'}{6} -3$ 
conditions are needed respectively for the anomaly-free solution.
In section 3 we give all gauge conditions eliminating ghosts in the
physical states of the composite string model and the spectral equations
for the quantum numbers.

\section{The critical composite superconformal
string model}

In order to describe the physical hadronic states with their 
quantum numbers and masses we have to consider a
string model in the four dimensional space-time.
Such models can be obtained by
compactification of the critical ten dimensional superstring. One 
way of compactification to four dimensions is the fermionization
of six dimensions \cite{abk}-\cite{abkw}, 
namely, by introducing free world sheet fermions 
$\nu(\sigma, \tau)$ carrying
all internal quantum numbers of the string 
\cite{abk}, \cite{vak}, \cite{pan}.  
A generalization and concrete realization of this approach 
has been achieved in the critical composite superconformal 
string model \cite{k}, \cite{pr}. This model
unifies the superconformal structures of the Neveu-Schwarz operator 
$G_r\sp{(NS)}$ 
\cite{ns} and of the fermion operator $G_r\sp{(f)}$ \cite{vak} by 
introducing the composite superconformal operator
\begin{equation}\label{Gr}G_r=G_r\sp{(NS)}+G_r\sp{(f)}. \end{equation}
The operator $G_r\sp{(f)}$ is constructed in such a way, that
$G_r$ is a {\it singlet} in all quantum numbers, and
$$\{G_r\sp{(NS)},G_r\sp{(f)}\}=0.$$
The specific superconformal algebras of 
$G_r\sp{(NS)}$ and $G_r\sp{(f)}$ are closed separately
(see the Appendix).

The canonical superconformal operator $G_r\sp{(NS)}$ reads
\cite{bh},\cite{ns} 
\[G_r\sp{(NS)}=\frac{1}{2\pi i} \oint H^{\mu}(\p_\tau X_{\mu}) 
e^{+ir\tau} d\tau.\]
We use the notations
\[X_\mu^{(i)}=x_{0 \mu}^{(i)} + ip_{\mu}^{(i)} \ln z_i + 
\sum_{n\neq 0}\frac{a^{(i)}\sb{n\mu}}{i n}z_i^n, \,\,\, 
z_i= e^{-i\tau_i}\]
$$[a^{(i)}_{n\mu}, a^{(i)}_{m\nu}]= -n \, g_{\mu \nu} \delta_{n, -m},
\ \ \ g_{00}=1, \,\,\, g_{ii}=-1,$$
for the $i$-th string space-time coordinate of zero conformal weight,
and 
$$H_{\mu}^{(i)}(z_i)=
b_{r\mu}^{(i)}z_i^r,
\,\,\,\{b_{r\mu}^{(i)}, b_{s\nu}^{(i)}\}=-g_{\mu \nu} \delta_{r, -s}$$
for its superpartner in the Neveu-Schwarz string model. The 
NS-states are given by the Fock space of 
the products of the creation operators
$$\prod_{n,\mu}\prod_{m,\nu}
\{a^{\dagger}_{n\mu}\}\{b^{\dagger}_{m\nu}\}|0>\ .$$

As usual the operator $G_r$ of the composite string
shall be of the conformal weight $I_c=3/2.$ Therefore the
fermion operator $G_r\sp{(f)}$ is of the same conformal weight.
It can
be constructed as a three-linear combination \cite{abk} 
of the fermion fields $\nu_A $ of $I_c=$  1/2 and their currents
$J\sp{\nu_A}$ of $I_c=$ 1, i.e. \cite{vak}
\begin{equation}\label{grnu}G_r\sp{(f)}(\nu)=\sum_{A,B,C}
\nu_A\nu_B\nu_C\varepsilon\sb{ABC} +\sum_{A} J\sp{\nu_A}\nu_A\ .
\end{equation}
By construction
$G_r^{(f)}$ is a singlet in the quantum numbers. The fermion operator 
generates all internal quantum numbers, what we show in detail in the
next section. 

We turn now to the description of the structure of the fields $\nu_A$ 
and their currents $J\sp{\nu_A},$ entering in (\ref{grnu}). Let
$\psi_{\alpha}\equiv\psi\sb{\mu j},\ \mu=0,1,2,3,\ j=1,2$ 
be Majorana spinor with $I_c=1/2.$ Its eight components are
Lorentz spinors in $\mu$ and isospinors in $j$ 
simultaneously. The respective currents $J\sp{\nu_A}$
$$J\sp{\nu_A}
=  \tilde{\psi}_{\alpha} T_{{\alpha} \beta}\psi_{\beta},
\ \ \ \ \ \tilde{\psi}_{\alpha} =T_0 {\psi}_{\alpha} 
= \gamma_0 \tau_2 {\psi}_{\alpha},
$$
are of conformal weight one. These currents are non-zero, if
the respective 
$8 \times 8$ matrices  $T_0 T_{{\alpha} \beta}$ 
are  antisymmetric. The matrices $T_{{\alpha} \beta}$ can
be choosen of the form
$$\gamma_{\mu},\,\,\gamma_5\gamma_{\mu} \tau_i,\,\, \tau_i
, \,\,\,
\gamma_5 \tau_i,\,\,[\gamma_{\mu}, \gamma_{\nu}].$$
Thus we obtain the 28 components for the vector, axial vector, scalar, 
pseudoscalar and tensor currents 
$$J\sp{\nu_A}=  \tilde{\psi}_{\alpha} 
T_{{\alpha} \beta}\psi_{\beta}=\{J^V_{\mu},  J^A_{\mu i},  
J^S_{i},  J^P_{i},  J^T_{\mu \nu}\}.$$
They generate a Kac-Moody algebra, see \cite{kaku}.
With these currents we associate the respective
28 fermion components
\begin{equation}\label{nuA} 
\nu_A:= \{\psi\sb{\alpha}; \phi\sb{\mu},
\rho\sb{\mu i},\theta_i,\eta_i,\xi\sb{\mu\nu} \}\ ,
\ \ \ \mu=0,1,2,3,\ \ i=1,2,3.
\end{equation}
These fields satisfy the standard anticommutation relations.
We consider the physical quarks as compositions of the
elementary field components (\ref{nuA}).

We see, that the eight spinor components $\psi_{\alpha}$ 
generate 28 vector, axial vector, scalar, pseudoscalar and 
tensor fermionic components. 
Thus $G_r^{(f)}$ is a scalar in  $D'$=36 fermionic components. 

It is well known, that the Neveu-Schwarz string is
critical in the space-time dimension $D=10.$ This comes from
the condition of cancellation of the conformal anomaly, or
equivalently
the nilpotency of the BRST-charge $\Omega^2=0$ \cite{abkw},
\cite{kaku}
\beq{*}
\frac{3}{2} D - 26 +11 = \frac{3}{2} (D-10)=0.
\eeq
{\it We point out, that
the new composite string remains critical for D=10.}
This is due to the ghost contributions 
{\footnote{The
ghost contribution is given by $c_g=-2\epsilon (6I_c(I_c-1) +1),$ 
where
$\epsilon=1$ is the ghost statistic for integer $I_c$ 
and $\epsilon=-1$ is the ghost statistic for half integer $I_c.$ 
Hence
$N$ fermions with $I_c=1/2$ contribute $c_g=-N$ 
and $N$ corresponding currents with $I_c=1$ contribute $c_g=-2N.$ }} 
from the introduced fermionic fields and their currents \cite{k}.
Indeed, in the composite model the condition (\ref{*}) turns into
\beq{nilp}
\frac{3}{2} D - 26 +11 +(\frac{1}{2} D' -3N) = 
\frac{3}{2}(D-10)+(\frac{1}{2} D' -3N) =0,
\eeq
where $-3N$ denotes the ghost contribution.
In the dimension $D=10$ the condition
(\ref{nilp}) is satisfied, if we choose $N=D'/6$ \cite{k}.

On the other hand,
compactifying the string to the space-time dimension
$D=4,$ we can impose $N=\frac{D'}{6}-3$ gauge constraints. 
 These gauge constraints on the fermionic sector 
are used to cancel the superfluous (in compairison with the 
Regge spectrum) states, and to shift the vacuum to $G_{-1/2}|0>.$ 
This shift ensures the absence of a tachyon in the spectrum. 
 
The amplitude of the hadron interactions are constructed by the 
operator formalism.
The quantum numbers spin, isospin, electric charge, hyper-charge,
baryon charge, strangeness, charm, beauty and top and the masses 
of the hadronic states are included in the structure of
the hadronic wave function which enters in the vertex operator 
structure, and hence in the structure of the N hadrons amplitude 
$A_N.$

%%The vertex operator is defined in the interaction representation as
%
%%$$V_i(z_i) = z_i^{-L_0} [G_r, W_i(z_i=1)] z_i^{L_0},$$
%$I_c=1$
%%where $W_i(z_i)=z_i^{-L_0}  W_i(z_i=1) z_i^{L_0}$ and $W_i(1)$ is the 
%%physical wave function at the time point of interaction $\tau_i=0$ 
%%(or $z_i=e^{-i \tau_i}=1$ in $z$ representation). 
The factorization of the amplitude gives the wave function of the 
corresponding hadronic state. 
This wave function contains a certain  number of 
components, corresponding to its experimental mass.
These components carry all hadron quantum numbers and will reproduce 
explicitly 
the quantum numbers in the hadronic amplitude
structure.
This has been shown in \cite{J58} where the
superconformal string amplitudes for $\pi$ mesons interactions 
have been constructed
and in \cite{pan}, \cite{pr} where the general $N$ hadrons interaction 
amplitudes are obtained as multiparticle generalization of the
Lovelace-Shapiro amplitude and are treated as composite
superconformal strings. 

Duality, crossing and cyclic symmetry for these 
superconformal composite string amplitudes hold together
with the description of hadron quantum numbers.

\section{The Quantum Numbers}

In this section we give a possible explicit construction of the quantum 
numbers of hadron wave functions. This wave functions shall
satisfy a set of gauge constraints
\begin{equation}\label{grph}
L_n|\mbox{phys}>=0,\ \ n>0,
\ \ \ \ G_r|\mbox{phys}>=0,\ \ r>0,
\end{equation}
which eliminate ghosts in the physical states. 
Here $L_n=L_n^{(NS)}+L_n^{(f)}$ 
denotes the Virasoro operator, the fermion part
of which is explicitly given in the Appendix.
As usual we impose the mass shell condition
\beq{msh}
(L_0-1/2)|\mbox{phys}>=0,
\eeq
where the intercept $1/2$ is fixed by the 
nilpotency of the BRST charge.

In addition to this
we have to require new gauge constraints 
\beq{gc}
\nu_r^{(l)}|\mbox{phys}>=0,
\ \ \ \ J^{\nu_r^{(l)}}|\mbox{phys}>=0, \,\,\,l=1,...,
N=\frac{D'}{6}-3,
\eeq
eliminating  
the ghosts contributions from the added fermionic fields and
their currents. 
Such new constraints are in agreement with the anomaly 
cancellation condition (\ref{nilp}).
They are responsible for the 
reduction of the composite string spectrum 
to the Regge resonance spectrum.
The gauge constraints (\ref{grph})-(\ref{gc}) build
a supermultiplet $L_n, G_r, J^{\nu_r^{(l)}}, \nu_r^{(l)}$
with the
conformal weights $I_c=2,\frac{3}{2},1,\frac{1}{2}$ respectively.

We search for a common system of
eigenfunctions for (\ref{grph})-(\ref{gc}) 
and the spectral equations
\beq{no}
{\hat{{\cal Q}}}_m|\mbox{phys}>= q_m  |\mbox{phys}>
\eeq
on the remaining $D' - N$ fermion components. 
The operator $\hat{{\cal Q}}_m$  runs over the quantum number 
operators for the
spin, isospin, electric charge,
hypercharge, baryon charge, strangeness, charm,
beauty and top.

An appropriate choice of the quantum number operators
are those zero components of the 
corresponding currents 
\begin{equation}\label{tok} 
({\cal{J}}\sp{\nu_A})_0=T\sp{\nu_A}:=\{G_r\sp{(f)},(\nu_A)\sb{-r},\}
\end{equation}
which are Number operators, i.e. count the component carrying the spin,
isospin, baryon charge and so on. 

The operators (\ref{tok}) automatically commute with the Virasoro 
operators $L_n\sp{(f)}$ and  $G_r\sp{(f)}$
$$[L_n\sp{(f)},T\sp{\nu_A}]=0,\ \ \ [G_r\sp{(f)},T\sp{\nu_A}]=0,$$
what ensures the existence of a common system of eigenfunctions.

Since $G_r\sp{(f)}$ is a quantum number singlet, 
the charge $T\sp{\nu_A}$ carries the same quantum numbers as
the field $\nu_A.$
Therefore the isospin operator shall be defined as 
\begin{equation}\label{Tie} T_i^\eta:=
\{G_r\sp{(f)},(\eta_i)\sb{-r}\}.
\end{equation}
It generates the algebra 
$[T_i^\eta ,T_j^\eta ]=\varepsilon\sb{ijk}T_k^\eta$ 
and has eigenvalues 
$$(\hat{T}^\eta_i)^2|\mbox{phys}>=T^\eta_i(T^\eta_i+1)|\mbox{phys}>.$$

Analogously the Lorentz spin operator is defined by
\beq{J}
J\equiv (k_\mu T\sp{\xi\sp{\mu\nu}}):=
\{G_r\sp{(f)},(k_\mu \xi\sb{\mu\nu})\sb{-r}\}
\eeq
with the momentum $k_\mu.$ The specific realization of the spin and 
isospin operators is given in the Appendix.

The baryon charge $B=1$ is carried by hadron states with half-integer 
Lorentz 
spin. Among the fields in (\ref{nuA})  only $\psi$ is a Lorentz spinor. 
Therefore, if a wave function contains an even number of $\psi$ 
components, it is a meson, otherwise it is a baryon. Thus the  baryon 
charge is defined by 
\beq{A}
B=\frac{1}{2}(1-(-1)\sp{N_B})
\eeq
where $N_B =\sum_{l}\tilde{\psi}\sb{-l}\psi_l$ is the number of spinor 
$\psi$-fields.

Among the currents (\ref{tok}) one can not find other
operators with Number structure. 
On the other hand all 36 components (\ref{nuA}) were used in the
specific realization of (\ref{Tie})-(\ref{A}). 
Therefore with them we can only construct meson and baryon states 
containing $u$ and $d$ quarks.
To define  the quantum numbers strangeness $\hat{s}$, 
charm $\hat{c}$, beauty $\hat{b}$ and top $\hat{t}$ independently, 
we have to extend the set (\ref{nuA}) of 36 fermion components 
with an analogous partner-set of 36 field components 
(Lorentz scalars and isoscalars) 
\begin{equation}\label{omA}
\omega_A:=\{\chi_\alpha;\,\sigma_i,\sigma\sb{jk}\}\ ,
\ \ \ \alpha=1...8,\ i,j,k=1...7.
\end{equation}
The fermion fields (\ref{omA}) show the 
following anticommutation properties 
\[\{(\tilde{\chi}_\alpha)_r,(\chi_\beta)_s\}=
\delta\sb{\alpha\beta}\delta\sb{r,-s},
\ \ \ \ \tilde{\chi}_\alpha = {\chi}_\alpha\gamma_0\tau_2,\]  
\[\{(\sigma_i)_r,(\sigma_j)_s\}=\delta\sb{ij}\delta\sb{r,-s},\]
\[\{(\sigma\sb{ij})_r,(\sigma\sb{kl})_s\}=(\delta\sb{ik}\delta\sb{jl}-
\delta\sb{il}\delta\sb{jk})\delta\sb{r,-s},
\ \ \ \sigma\sb{ij}=-\sigma\sb{ji}.\]
From the 8 component spinor $\chi_\alpha$ we can define 28 
components of the Kac-Moody currents
\beq{Jw}
J^{\omega_A}=\Bigl\{J_i:=\tilde{\chi}\Gamma_i\chi ;\ \ \ 
J\sb{jk}:=\tilde{\chi}[\Gamma_i,\Gamma_j]\chi\Bigr\}.
\eeq
which are Lorentz scalars and isoscalars.
Here $\Gamma_i $ denote the $8\times 8$  Clifford matrices, 
obeying the standard anticommutation relations 
$\{\Gamma_j,\Gamma_k\}=2\delta\sb{jk}.$

The currents $J_i, J\sb{jk}$ carry the same quantum numbers as
the fields $\sigma_i, \sigma\sb{jk}.$ 
This guaranties the scalarity of the superconformal operator
\begin{equation}\label{Grw} G_r\sp{(f)}(\omega)=\sum_{ABC} 
\omega_A\omega_B\omega_C\epsilon\sb{ABC}+
\sum_{A}J\sp{\omega_A}\omega_A.\end{equation}
Composing the superconformal operators (\ref{grnu}) and (\ref{Grw})
\begin{equation}\label{GFR} 
G_r\sp{(f)}=G_r\sp{(f)}(\nu)+G_r\sp{(f)}(\omega)\end{equation} 
we describe the full set of hadron quantum numbers
{\footnote{$G_r\sp{(f)}$ is a quantum number singlet in $D'=72$, which 
can be reduced to D=10 according to (\ref{nilp}) by imposing 
$N=D'/6=12$ conditions on the physical states, 
i.e. 6 on each of the sets (\ref{nuA}) and (\ref{omA}).
Then for the compactification of the composite string to D=4
are used $N=D'/6-3=9$ constraints.}}.

From (\ref{Tie}) the third isospin component in the $\nu_A$-space is 
given by
\[T_3^\eta :=\{G_r\sp{(f)},(\eta_3)\sb{-r}\}.\] 
By construction $\sigma\sb{12}\in\omega_A$ 
is the partner-component  to $\eta_3.$ 
Therefore we define the isospin projection in $\omega_A$  by
\beq{toka}
T_{12}:=\{G_r\sp{(f)},(\sigma\sb{12})\sb{-r}\}.
\eeq
The full projection of the isospin of the hadron state corresponds to 
the sum
\begin{equation}\label{That}
\hat{T}_3|\mbox{phys}>=(T_3^\eta+T\sb{12})|\mbox{phys}>\ .
\end{equation}

In order to describe the flavour quantum numbers strangeness, charm,
beauty  and top we have to construct four commutating operators in 
$\omega_A$
with structure similar to the zero component of the currents 
$J^{\omega_A}.$
We will find them using the 
properties of the spinor representation of the $O(6)$ group. 
The 8 component spinors in the  compactified
$D=6$ space transforms under $O(6).$ 
The generators of this group are given via the $8\times 8\ \Gamma_i$ 
matrices
by $[\Gamma_i,\Gamma_j]= 4i M_{ij},\,\,\, i,j=1...6.$ 
It holds $[M_{ij},M_{kl}]=0,$ if and only if $i\neq k,l$ and 
$j\neq k,l.$
Thus among the generators one can choose 
not more then three commutating with each other $M_{ij},$
e.g. $M_{12}, M_{34}, M_{56}.$ Since
the Clifford matrices $\Gamma_i$ transforms under  $O(6)$  as vectors 
$$[M_{ij},\Gamma_k] = i(\delta_{ik}\Gamma_j-\delta_{jk}\Gamma_i),
\ \ i,j,k=1...6,$$
neither of the matrices $\Gamma_1,...,\Gamma_6$ commutes with all of the
three generators $M_{12}, M_{34}, M_{56}.$ On the other hand, the matrix
$\Gamma_7 := \prod_{i=1}^6 \Gamma_i$ commutes with all generators 
$M_{ij}.$ Thus $\Gamma_7$ is the fourth independent operator.

According to this we introduce first three independent generators 
$T_{12}, T_{34},$ $T_{56}.$
We choose $T_{12}$ according to (\ref{toka}),
since it defines the $s,c,b,t$-flavour part of the isospin projection.
The remaining two generators are given by 
\begin{eqnarray} 
\label{tokb} T\sb{34}:=\{G_r\sp{(f)},(\sigma\sb{34})\sb{-r}\}\\
\label{tokc} T\sb{56}:=\{G_r\sp{(f)},(\sigma\sb{56})\sb{-r}\}.
\end{eqnarray}
The fourth charge operator has the form
\begin{equation}\label{curr} 
T_7:=\{G_r\sp{(f)},(\sigma_7)\sb{-r}\}.
\end{equation}
It commutes with (\ref{toka}), (\ref{tokb}), (\ref{tokc}) 
and with $L_n, G_r.$ Hence there exists
a common system of eigenfunctions for all these operators.
 
The operators $T_{12}, T_{34}, T_{56}, T_7$ 
can not directly be identified as flavor quantum number operators.  
In fact one has to consider the following four linear combinations,
which define strangeness, charm, beauty and top
\beq{[s}
\hat{s}:=-\frac{1}{2}(T\sb{34}+T\sb{56})-\frac{1}{2}(T\sb{12}+T_7),
\eeq
\beq{[c}
\hat{c}
:=-\frac{1}{2}(T\sb{34}+T\sb{56})+\frac{1}{2}(T\sb{12}+T_7),
\eeq
\beq{[b}
\hat{b}:=-\frac{1}{2}(T\sb{12}+T\sb{34})+\frac{1}{2}(T\sb{56}+T_7),
\eeq
\beq{[t}
\hat{t}:=+\frac{1}{2}(T\sb{12}+T\sb{56})-\frac{1}{2}(T\sb{34}+T_7).
\eeq
Thus the physical quarks are described
as superpositions of elementary string fields. Hence the quarks 
are composite objects in this model. 

If we substitute (\ref{That}) and (\ref{[s})-(\ref{[t}) into
the expression for the electric charge 
$$Q=T_3+(\hat{s}+\hat{b}-\hat{c}-\hat{t}+B)/2$$ 
with the baryon charge $B,$
we obtain
\begin{equation}\label{Q} Q=T^\eta_3+B/2.\end{equation}
This can be taken as a definition, which is 
independent of the quark flavors contained in the hadronic state. 
Analogously 
the expression for the hypercharge in the quark model 
$$Y=\hat{s}+\hat{b}-\hat{c}-\hat{t}+B$$
together with (\ref{[s})-(\ref{[t}) gives
\begin{equation}\label{Y} Y=B-2T\sb{12}.\end{equation}
In case of $T\sb{12}=0$ we have  $Y=B$ for all states, containing 
only $u$- and $d$-quarks.

The above constructions give a
realization of the hadron quantum numbers 
$Q,B,T_3,\hat{s},\hat{b},\hat{c},\hat{t}$ 
in terms of the composite string operators.
Each given hadron state can be interpreted as a
wave function with the respective 
quantum numbers in the composite string model.
 
The spin, isospin and baryon charge in the first generation of 
({\tiny$\begin{array}{cc}u\\d\end{array}$})-quarks 
are realized from the $\nu_A$-set, 
while the $s,c,b,t$ flavours 
are realized on the partner set $\omega_A.$ 
Thus the respective wave functions have the structure 
\[<\mbox{meson}|=<0,k|F_{\nu_A}F_{\omega_A},
\ \ <\mbox{baryon}|=<0,k|\psi F_{\nu_A}F_{\omega_A}\]
with $<0,k|=<0|\exp (ikx).$ In this $F_{\nu_A}$
is a compositions of components from (\ref{nuA}), 
symmetric or antisymmetric with respect to the spin and isospin, while 
$F_{\omega_A}$ carries the quantum numbers 
$\hat{s}, \hat{c}, \hat{b}, \hat{t}$, and the  isospin projection
$T\sb{12},$ and contains only components from $\omega_A.$

The fermion superconformal operator $G_r\sp{(f)}$ has 
$D'=72$ internal degrees of freedom.
Notice, that the quark model contains  
6 flavors, 3 colours and 4 spin components, what also gives
total number of 72 degrees of freedom.

Finally we turn to the realization of the mass-spectrum of the hadron 
resonances. 
Let the strange fermion component gives a contribution of 
$\Delta M^2 = 0.3\,\, \mbox{GeV}^2$ to the mass of the physical state.
This agrees with the experimental fact of parallel $\rho$-meson- and
$K^*$-trajectories
\[M^2_{K^*} - M_\rho^2 = \frac{1}{2\alpha '} = 0.3 \mbox{GeV}^2.\]
In this case the composite string model gives a much better description 
of the physical mass spectrum then the quark model. In the quark model 
these trajectories are not parallel.

\section{Discussion}

A description of the hadron quantum numbers has been given
in the composite string model in the simplest case
of 8 component spinors in a compactified six-dimensional space. Another 
possibility is to consider 32 component
spinors, what is natural in the critical D=10 string. There the number
of currents obtained as pairs of 32 component spinors is equal to 496. This
gives $D'=528$ fermionic degrees of freedom. Here we impose 
$N=D'/6=88$ gauge conditions on physical states. 
This allows a simultaneous description of the quarks and leptons
in terms of the fermionic sector of the model.
This is a subject of further publications.

\section{Appendix}

We give here the concrete realization of the described operators. 
In \cite{vak} the
fermion 
superconformal operator $G_r^{(f)}(\nu)$ was
introduced as
\[G_r^{(f)}(\nu) = \frac{1}{\sqrt{7}}\{\frac{1}{4i}
(\tilde{\psi}\hat{\phi}\psi)  
 + \frac{1}{4}(\tilde{\psi}\gamma_5\tau_i\hat{\rho}\psi) + 
\frac{1}{4}(\tilde{\psi}\tau_i\psi)\eta_i + 
\frac{1}{4}(\tilde{\psi}\gamma_5\tau_i\psi)\theta_i +\]
\[ + \frac{1}{8i}(\xi^{\mu\nu}\tilde{\psi}\frac{1}{2}[\gamma_\mu 
,\gamma_\nu]\psi) + \frac{1}{i}(\phi\rho_i)\theta_i + 
\frac{1}{2i}(\theta_i\theta_j\eta_k)\varepsilon_{ijk} - 
\frac{1}{2i}((\rho_i\rho_j)\eta_k)\varepsilon_{ijk} + \]
\[+ \frac{1}{6i}(\eta_i\eta_j\eta_k)\varepsilon_{ijk} - 
\frac{1}{2i}(\xi^{\mu\nu}\rho_{\mu i}\rho_{\nu i}) - 
\frac{1}{2i}(\xi^{\mu\nu}\phi_\mu\phi_\nu) + 
\frac{1}{6i}(\xi^{\mu\nu}\xi_{\nu\lambda}\xi^{\lambda}_{\mu})\}_r\] with 
\[(\eta_i\eta_j\eta_k)_r = \sum_{r_1+r_2+r_3=r} 
(\eta_i)_{r_1}(\eta_j)_{r_2}(\eta_k)_{r_3}.\] 

By analogy the
operator $G_r^{(f)}(\omega)$  is given by
\[G_r^{(f)}(\omega)=\frac{1}{\sqrt{7}}
\{(\tilde{\chi}\Gamma_i\chi\sigma_i)_r + 
(\tilde{\chi}[\Gamma_j,\Gamma_k]\chi\sigma_{jk})_r + 
(\sigma_i\sigma_j\sigma_{ij})_r + 
(\sigma_{ij}\sigma_{jk}\sigma_{ki})_r\},\] 
where the matrices $\Gamma_i$ are defined as 
\[\Gamma_i=(\Gamma_{i'},\Gamma_{\mu}),\ i'=1,2,3,\ \ \mu=4,5,6,7,\] 
\[\Gamma_{i'}=\gamma_5\otimes\tau_{i'},\ \ i'=1,2,3,\]
\[\Gamma_4=\gamma_1\otimes 1,\ 
\Gamma_5=\gamma_2\otimes 1,\ \Gamma_6=\gamma_3\otimes 1,\ 
\Gamma_7=\Gamma_1\Gamma_2\Gamma_3\Gamma_4\Gamma_5\Gamma_6.\]

The full fermion operator is given 
by (\ref{GFR}). Its superconformal algebra is closed
\[\{G_r^{(f)},G_s^{(f)}\} = 2 L^{(f)}_{r+s} + \frac{D'}{6}
(r^2-\frac{1}{4})\delta_{r,-s},\]
i.e. the superconformal structure of the model is conserved.

The Virasoro conformal algebra is satisfied
\[[L_n^{(f)},L_m^{(f)}] = (n-m)L_{n+m}^{(f)} + 
\frac{D'}{24}(n^3-n)\delta_{n,-m},\] 
and
the conformal and superconformal operators commutation relation hold
\[[L_n^{(f)},G_r^{(f)}] = (\frac{n}{2} -r)G_{n+r}^{(f)}.\] 
The factor $\frac{n}{2}-r$ implies, that $G_r$ is of
the conformal weight $3/2.$

The Virasoro operator 
$L_n^{(f)}$ correspond to the fermion part of the action of the free string
\[S=\frac{i}{8\pi \alpha '}\int d\tau d\sigma (\tilde{\psi}\hat{\partial}\psi 
- \phi\hat{\partial}\phi - \rho\hat{\partial}\rho - \eta\hat{\partial}\eta + 
\theta\hat{\partial}\theta + \xi\hat{\partial}\xi + \chi\hat{\partial}\chi + 
\sigma_i\hat{\partial}\sigma_i + \sigma_{ij}\hat{\partial}\sigma_{ij}).\]
Next we give the realization of the 
the quantum number operators 
in terms of the components of the elementary string fields. 
The isospin operator 
$T^\eta_i$ is given by 
\[T^\eta_i = 
\sum_{j,k}:(\eta_j\eta_k\varepsilon_{ijk})_0: + 
\frac{1}{2}\sum_{l>0}(\tilde{\psi}_{-l}\tau_i\psi_l) + 
\sum_{j,k}:(\theta_j\theta_k+\rho_{\mu j}\rho_{\mu k})_0\varepsilon_{ijk}:.\] 
The Lorentz-spin operator is defined as
\[(k_\mu T^\xi_{\mu\nu}) = -
\frac{1}{4} \sum_{l>0} k_\mu\tilde{\psi}_{-l}[\gamma_\mu,\gamma_\nu]\psi_l + 
\sum :(k_\mu\rho_{\mu i}\rho_{\nu i} + k_\mu\phi_\mu\phi_\nu + 
k_\mu\xi_{\mu\lambda}\xi_{\lambda\nu})_0:.\] 
The operators
$T_{12},T_{34},T_{56},T_7$ can be realized as 
\[T_{12} = \sum_{r>0} 
\tilde{\chi}_{-r}\tau_3\chi_r + \sum :(\sigma_1\sigma_2)_0: + \sum 
:(\sigma_{1k}\sigma_{2k})_0:,\] \[T_{34} = \sum_{r>0} \tilde{\chi}_{-
r}[\Gamma_3,\Gamma_4]\chi_r + \sum :(\sigma_3\sigma_4)_0: + \sum_{k} 
:(\sigma_{3k}\sigma_{4k})_0:,\] \[T_{56} = \sum_{r>0} \tilde{\chi}_{-
r}[\Gamma_5,\Gamma_6]\chi_r + \sum :(\sigma_5\sigma_6)_0: + \sum_{k} 
:(\sigma_{5k}\sigma_{6k})_0:,\] \[T_{7} = \sum_{r>0} \tilde{\chi}_{-
r}\Gamma_7\chi_r + \sum_{k} :(\sigma_{7k}\sigma_k)_0: .\]

\end{document}